\documentclass{elsart}
 \usepackage{graphics}
\usepackage{amssymb}

\newcommand\mRe{{\rm Re\,}}
\newcommand\mIm{{\rm Im\,}}
\begin{document}

\begin{frontmatter}



\title{Plasmon mechanism of light transmission through a metal film or a plasma layer}


\author{Yu.P. Bliokh}

\address{Department of Physics, Technion, 32000 Haifa, Israel}
\ead{bliokh@physics.technion.ac.il}
\begin{abstract}
It is shown that a smooth metal film (or a plasma layer) can be made transparent for an electromagnetic wave when two identical subwavelength diffraction gratings are placed on both sides of the film. The electromagnetic wave transmission through the metal film is caused by excitation of evanescent surface waves (plasmons) and their transformation into propagating waves at the gratings.  A model which is developed analytically shows that the problem of the wave transmission is physically equivalent to the problem of excitation of two coupled resonators of evanescent waves which are formed at the two film surfaces. 
\end{abstract}

\end{frontmatter}

\section{Introduction}

The anomalous transmission of  light through periodically perforated metal films has recently attracted enhanced attention due to the fundamental significance of this effect and to its numerous possible applications, e.~g. in optoelectronics, subwavelength lithography {\it etc}. It has been shown for the first time in Ref.~1 that a periodic set of subwavelength holes enhances at some frequencies the transparency of an optically thick film by several orders of magnitude. It was pointed out in this paper that this phenomenon can be related to resonance excitation of surface waves -- plasmons -- at the metal surface. These plasmons are trapped electromagnetic surface modes propagating along the interface between two media which have permittivities or permeabilities with opposite signs. The electromagnetic fields of the plasmons decay exponentially into both media. Plasmons were discovered for the first time at the interface between a vacuum and a plasma-like medium \cite{Ritchie}.

Plasmons can be excited at a smooth dielectric-metal  boundary by an incident electromagnetic wave when the wave is totally internally reflected at a near-boundary inhomogeneity of the dielectric  permittivity \cite{Otto,Sarid,Dragilla,Welford}. An artificial periodical (e.~g. corrugated or perforated) or natural accidental inhomogeneity of a metal boundary also allows the excitation of plasmons by an incident electromagnetic wave. Wood's paper \cite{Wood} of 1902 can be considered as the starting moment for investigations in this direction. In spite of such a long history this problem continues to attract attention (see, e.~g., \cite{Wedge,Darmanyan,Bonod,Giannattasio,Marquier}).

There have been many theoretical works which suggest models that try to clarify the physical mechanism that is responsible for the anomalous light transmission through the perforated metal film. Considerable part of these works is based on results of numerical simulations (see, e.~g., \cite{Bonod,Giannattasio,Popov2000,Popov2004,Martin-Moreno,Krishnan,Esteban,Lezec,Tan,Avrutsky,Qing,Sarrazin}) and only some of them propose analytical investigations \cite{Darmanyan,Dykhne,Kats,Genchev}. Most of the authors share the opinion that the enhanced transparency of the perforated metal film is accompanied by excitation of plasmons. However, various authors consider different mechanisms of coupling between the plasmons on both surfaces of the film, which is the key element for the electromagnetic wave transmission. In Ref.~\cite{Krishnan}, for example, the holes are considered as subwavelength cavities for evanescent waves coupling the plasmons on either side of the film. The theory developed in Ref.~\cite{Martin-Moreno}  considers coupling of the plasmons through the evanescent electromagnetic modes in the holes which are treated as subwavelength waveguides. In Ref.~\cite{Lezec} the enhanced transparency is explained by interference of diffracted evanescent waves.

Other authors consider the holes in the film as a periodical inhomogeneity of the metal surface \cite{Bonod,Tan,Avrutsky} or as a periodical modulation of the metal dielectric permittivity \cite{Dykhne,Darmanyan,Kats}. In these models plasmons on both sides of the film are coupled due to the overlapping of their fields in the metal body and periodical modulation of the surfaces or of the dielectric permittivity is necessary for coupling between incident and transmitted electromagnetic waves and plasmons. 

It is appropriate to mention here that there are also diametrically opposite points of view on the role of plasmons in the electromagnetic wave transmittance through the perforated (or corrugated) metal film. On the one hand, it is asserted in Ref.~\cite{Qing} that the excitation of plasmons  decreases the transparency of the film. On the other hand, it is shown in Ref.~\cite{Esteban} that excitation of surface waves, whatever its nature (spoof plasmons \cite{Pendry2} or Brewster-Zennek modes \cite{Sarrazin}, for example), and not only plasmons, leads to the increase of the film transparency. 

We do not set ourselves to clarify the contribution of one or another physical mechanism in the transparency enhancement as an object of this work. In this paper we do present a simple model which allows us to investigate analytically the plasmon mechanism of light transmission through an optically thick metal film. In order to exclude any other possible mechanism we consider a construction which is noticeably simple for analytical description and in which only the plasmon mechanism is possible. Instead of a {\it perforated} film, a {\it smooth} metal film with two diffraction grids placed on both sides of the film are considered. Besides the considerable simplification of the analytical description and the clear physical treatment of the electrodynamical properties of this ``sandwich'', such a construction may possess some technological merits.

The metal film is an open (not bounded by walls) resonator of surface waves. In the absence of dissipation, the resonator Q-factor is infinitely large. The resonator excitation by an external source results in a significant growth of an eigenmode amplitude. Since the eigenmode fields are concentrated near both sides of the film, the eigenmode amplitude growth leads to an increase of the fields  at the opposite side of the film with respect to the source position that can be regarded as an anomalous film transparency. The aim of this work is to show that such interpretation of the anomalous transparency is not only a convenient model, but also reflects adequately the physics of the process. The resonator properties -- eigenfrequencies and geometry of eigenmodes fields -- completely, qualitatively and quantitatively, determine the dependence of the film transparency  on the system parameters.

\section{How a metal film can be made transparent for light?}

Let us start from the simplest problem considering a plane monochromatic electromagnetic wave incidence at a vacuum-metal interface. The electromagnetic properties of the metal will be described by its dielectric permittivity (Drude model) $\varepsilon_p=1-\omega_p^2/\omega^2$, where $\omega_p$ and $\omega$ are the metal electron plasma frequency and the incident wave frequency, respectively. In this approximation there is no difference between a metal film and a plasma layer and we will use both these terms as synonyms. Let us assume that the $z$-axis is directed along the interface normal, the $x$-axis lies in the plane of incidence and the metal occupies a right-hand half-space. The incident wave fields are proportional to $\exp (ik_xx+ik_zz-i\omega
t)$, where $\vec{k}$ is the wave vector. The reflected and the refracted wave fields are proportional to $\exp (ik_xx-ik_zz-i\omega t)$ and $\exp
(ik_xx+ik_{pz}z-i\omega t)$ correspondingly. Here $k_{pz}=\sqrt{k_0^2\varepsilon_p-k_x^2}$ is the $z$-component of the wave vector in the metal and $k_0=\omega/c$. Dissipation processes in the metal will be neglected.

When the wave frequency $\omega$ is smaller than the plasma frequency $\omega<\omega_p$, then the $z$-component of the wave vector in the plasma, $k_{pz}$, is imaginary for an arbitrary value of  $k_{x}$, i.~e. the refracted wave decays exponentially deep into the plasma and its amplitude decreases as $\exp(-|k_{pz}|z)$. If the plasma occupies a whole half-space $z>0$, then the incident wave energy is completely transmitted  to the reflected wave energy. If the plasma layer thickness $h_p$, is a finite quantity, then the wave passes partially through the plasma but the transmitted wave amplitude is exponentially small, $\sim\exp(-|k_{pz}|h_p)$.

The plasma permittivity is negative, $\varepsilon_p<0$, when $\omega<\omega_p$. As is well known, along the interface between two media in which the dielectric permittivities have opposite signs, a surface electromagnetic wave can propagate. The surface wave fields decay exponentially on either side of the interface. The magnetic field of this wave is parallel to the interface plane. Let the $y$-axis be directed along the magnetic field of the wave. The dispersion relation for the surface wave propagating along the vacuum-plasma boundary can be written in the form:
\begin{equation}\label{eq1}
{\sqrt{k_0^2|\varepsilon_p|+k_x^2}\over|\varepsilon_p|\sqrt{k_x^2-k_0^2}
}=1.
\end{equation}
It is not difficult to show that Eq.~(\ref{eq1}) possesses a real solution $k_x(\omega)$ if $\omega^2<\omega_p^2/2$. This solution corresponds to an undamped wave -- surface plasmon -- propagating along the interface.

The surface wave phase velocity $\omega/k_x$ is directed along the interface and is less than the speed of light $c$. As a result, this wave cannot be excited by an incident {\it propagating} electromagnetic wave. The surface wave can be excited by an ``incident'' {\it nonpropagating} (evanescent) wave for which $k_x^2>k_0^2$. In this case the term ``incident'' means the wave whose amplitude decays exponentially as $\sim\exp(-\sqrt{k_x^2-k_0^2}z)$ on approaching the interface from a source placed in the half-space $z<0$.

Let us assume that the magnetic field of the evanescent incident wave is directed along the $y$-axis and is equal to $H_1\exp(-kz)$, and the magnetic fields of the surface wave on either side of the interface, in the $z<0$ and $z>0$ half-spaces, are equal to $H_2\exp(+kz)$ and $H_3\exp(-k_pz)$, respectively. Here $k=\sqrt{k_x^2-k_0^2}$,
$k_p=\sqrt{k_x^2+|\varepsilon_p|k_0^2}$ and a common factor  $\exp(ik_xx-i\omega t)$ is omitted. Using the standard boundary conditions it is easy to express the amplitudes $H_2$ and $H_3$ in terms of $H_1$:
\begin{equation}\label{eq2}
H_2=H_1{1+k_p/(k|\varepsilon_p|)\over
1-k_p/(k|\varepsilon_p|)},\hspace{0.2cm}H_3=H_1{2\over 1-k_p/
(k|\varepsilon_p|)}.
\end{equation}

Equations (\ref{eq2}) contain a resonance denominator whose zero coincides with a root of the dispersion equation (\ref{eq1}). The physical meaning of these formulae is very simple: the interface forms an original resonator whose eigenmode is the surface wave. In the model under consideration (dissipationless and unbounded in the $x$ and $y$ directions) the resonator energy loss is equal to zero. The incident wave $H_1$ plays the role of an external force (a pump mode), therefore the amplitude of oscillations in the resonator increases without limit when the incident wave frequency approaches the resonator eigenfrequency. 

When plasma occupies the space between two planes $z=0$ and $z=h_p$, then the second surface forms the same resonator. The resonators on either sides of the plasma layer are connected by fields which decay exponentially inward the plasma body. Therefore the resonators coupling coefficient is small (it is assumed that $k_ph_p\gg1$). A coupling of two identical resonators leads to eigenfrequency splitting into two frequencies which are shifted proportionally to the coupling coefficient. These eigenfrequencies correspond to symmetric and antisymmetric eigenmodes relative to the layer center.

Such a binary resonator can be excited in the same way as the single one, by an external nonpropagating electromagnetic wave. Solving the problem of such a wave passing through the plasma layer, one can obtain the expression for the magnetic field amplitude $H_{tr}$ of the transmitted wave on the other side of the layer, in the region $z>h_p$:
\begin{equation}\label{eq3}
H_{tr}={k|\varepsilon_p|\over k_p}{e^{-kh_p}\over D_0}H_1,
\end{equation}
where
\begin{equation}\label{eq4}
D_0={1\over4}\left(1-{k|\varepsilon_p|\over
k_p}\right)^2e^{k_ph_p}-{1\over4}\left(1+{k|\varepsilon_p|\over
k_p}\right)^2e^{-k_ph_p}.
\end{equation}

The equation $D_0=0$ is the dispersion equation whose roots determine the eigenfrequencies of the coupled resonators. Thus, in this case too the wave field amplitude approaches infinity when the pump wave frequency reaches one of the two eigenfrequencies of the binary resonator.

If it were not a conditional application of ``incident wave'' and ``transmitted wave'' conceptions to the evanescent waves, it would be possible to interpret expression (\ref{eq3}) not only as wave transmission through the plasma layer, but also as wave ``amplification'' by the layer. The ``perfect lens'' effect \cite{Pendry} is accounted for exactly by such ``amplification'' of evanescent waves by a left-handed material slab \cite{Ruppin,Liu,Fang,Rao,Bliokh}. 

In order to obtain real transmission of a propagating wave through the plasma layer it is necessary to transform (partially or completely) this wave into an evanescent one and to make an inverse transformation behind the layer. Since, as it follows from Eq.~(\ref{eq3}), the evanescent wave amplitude can be very large, essential enhancement of the wave transmission is expected even if the  transformation coefficient is small.

A one- or two-dimensional diffraction grating  is a simple ``transformer'' of propagating and evanescent waves one into another. A wave incident at the grating gives rise to a set of reflected and refracted waves whose wave vectors projections $\vec{k}_{\perp n,m}$ at the grating plane differ from the projection $\vec{k}_\perp$ of the wave vector of the incident wave by an integer number of grating vectors $\vec{k}_{g_{1,2}}$: $\vec{k}_{\perp
n,m}=\vec{k}_\perp+n\vec{k}_{g_1}+m\vec{k}_{g_2}$,
$n,m=0,\pm1,\pm2,\ldots$.

Secondary waves for which $k_{\perp n,m}<k_0$, are propagating waves, and waves for which $k_{\perp n,m}>k_0$ are evanescent waves. The last ones are localized near the grating plane and decay exponentially, moving away from the grating. Independently of the incident wave polarization there are, in the general case, such evanescent waves whose magnetic field have non-zero projection onto the grating plane. These waves can excite the surface wave resonator.

Thus it is expected that the plasma layer transparency for an electromagnetic wave, whose frequency is smaller than the plasma frequency, can be enhanced essentially when on each side of the layer  two diffraction gratings are placed. It will be shown below that the transparency of such a ``sandwich''  reaches 100\%.

\section{The diffraction grating model}

In the general case only one of the evanescent waves scattered on the grating can be used for the resonator excitation. The contribution of all other non-resonant waves in the interaction of the incident wave with the system under consideration is small and will be neglected below. Therefore, without loss of generality, we may consider the grating to be one-dimensional and regard the magnetic fields of the incident and scattered waves as lying in the grating plane and being perpendicular to the grating inverse wave vector.

The following dimensionless variables will be used below: $\xi=x\omega_p/c$, $\zeta=z\omega_p/c$, $\Omega=\omega/\omega_p$. The wave numbers (real or imaginary) of wave spatial harmonics will be normalized to $\omega_p/c$ and designated in the vacuum and in the plasma as $q$ and $\kappa$, respectively.

\begin{figure}[tbh]
\centering \scalebox{0.6}{\includegraphics{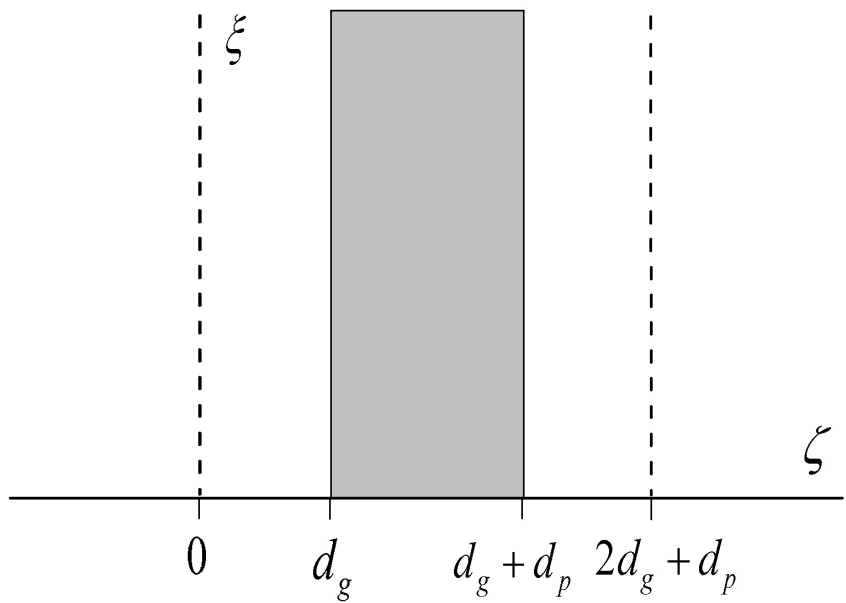}}
\caption{\label{Fig1}Plasma layer between two diffraction gratings.}
\end{figure}

Suppose that on each side of a plasma layer, which is situated between the planes $\zeta=d_g$ and $\zeta=d_g+d_p$, two diffraction gratings are placed in the planes $\zeta=0$ and $\zeta=2d_g+d_p$ (see Fig.~\ref{Fig1}). In order to avoid cumbersome expressions we will assume that the gratings are identical and placed at the same distances $d_g$ from the plasma boundaries, and the grating wave vector $\vec{q}_g$ is directed along the $\xi$ axis. 

Let us consider a plane electromagnetic wave passing through the grating-plasma-grating (GPG) system. It is assumed that the wave falls from left to right on the GPG system, the wave vector of the incident wave lies in the $\xi,\zeta$ plane and the wave magnetic field is parallel to the gratings planes.

The problem of a transmission coefficient calculation for a system with one or another shape of grating lies out of the framework of this paper. Therefore we will use the following simple model of wave scattering at a diffraction grating. 
Let us consider a modified wave equation for the wave magnetic field:
\begin{equation}\label{eq5}
{\partial^2 H\over \partial \zeta^2}+{\partial^2 H\over \partial
\xi^2}+\Omega^2
H+\delta(\zeta)\left(\mu_0+2\mu_1\cos
k_g\xi\right)H=0.
\end{equation}

Eq.~(\ref{eq5}) describes the wave passing through a $\delta$-shaped layer situated in the $\zeta=0$ plane. The layer permeability is varied harmonically in the $\xi$-direction. Such a layer posesses all the properties of a diffraction grating and therefore will be used for the description of waves scattering at the grating. The harmonic rule of the permeability variation does not restrict the model generality, because,  as it was mentioned above, the anomalous transparency of the plasma layer is connected with the excitation of a surface wave which is resonant with one of the spatial harmonics of the wave passing through the grating. Therefore, only one harmonic of the permeability variation plays a key role. The contribution of all the other harmonics is negligibly small and will be neglected further.

The parameters $\mu_0$ and $\mu_1$ describe different properties of the grating. The parameter $\mu_0$ determines the mean transparency of the grating. The parameter $\mu_1$ determines the coefficient referred above  of the mutual transformation of the propagating and evanescent spatial harmonics of the waves. Since the last parameter plays the key role, we can suppose at first that $\mu_0=0$. In accordance with Flouqet theorem let us look for Eq.~(\ref{eq5}) solution in the form:
\begin{equation}\label{eq6}
H(\zeta,\xi)=e^{iq_\xi\xi}\sum_nH_n(\zeta)e^{ink_g\xi}.
\end{equation}

A boundary condition for the expansion coefficients $H(\zeta)$ at the $\delta$-layer follows from Eq.~(\ref{eq6}):
\begin{eqnarray}\label{eq7}
H_n(-0)=H_n(+0),\nonumber\\
\left.{dH_n\over
d\zeta}\right|_{+0}-\left.{dH_n\over
d\zeta}\right|_{-0}+\mu_1\left[H_{n-1}(0)+H_{n+1}(0)\right]=0.
\end{eqnarray}

The conditions (\ref{eq7}) and standard boundary conditions at the plasma surface are enough for the problem solution.

\section{Electromagnetic wave transmission through the GPG system: A wave theory}

Let the incident wave frequency be $\Omega$. The wave magnetic field has the form $H_{1,0}e^{iq_\xi\xi+iq_0\zeta}$, where $q_0=\sqrt{\Omega^2-q_\xi^2}$. For the sake of simplicity let us assume that the grating period is less than the wavelength, i.~e. all the spatial harmonics with  $n\neq0$ in Eq.~(\ref{eq6}) are evanescent. Only two harmonics are of interest for the problem: the zero harmonic because the total energy flux behind the second grating is concentrated only in it, and one from the evanescent harmonics with $n\neq 0$ (let it will be $n=+1$ for definiteness) which can play the pump mode role for the surface waves resonator. The contribution of all the other harmonics will be neglected below.

\begin{figure}[tbh]
\centering \scalebox{0.6}{\includegraphics{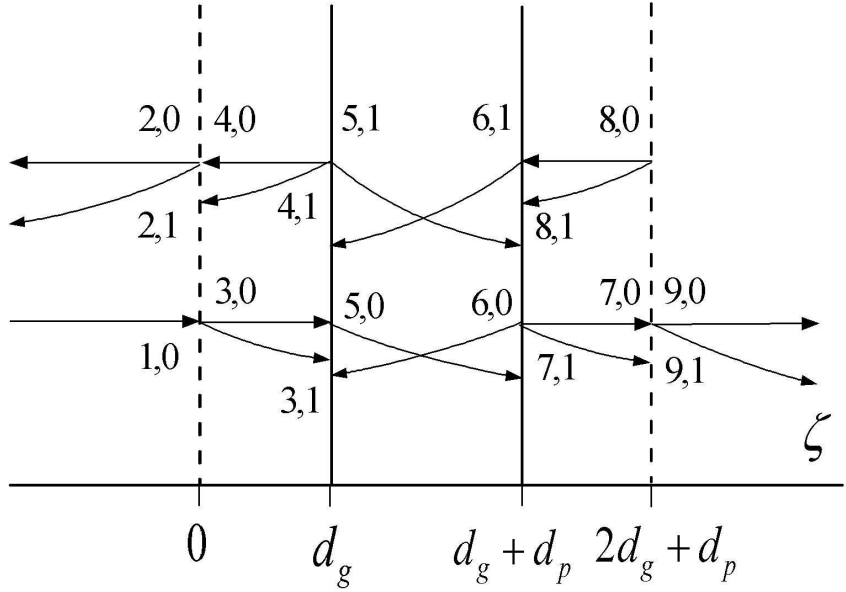}}
\caption{\label{Fig2}System of waves and their harmonics.}
\end{figure}

The system of waves and their harmonics which have to be taken into account for the transmission coefficient calculation, is shown in Fig.~\ref{Fig2}. The harmonic amplitudes will be marked by two indices. The first one corresponds to the wave number and the second one corresponds to the harmonic number. The propagation direction of the corresponding wave is marked by an arrow. As the ``propagation direction'' for the evanescent wave, the direction of its amplitude decrease is chosen. The propagating waves are represented by horizontal lines and the evanescent waves are represented by conventional curves which fall along the direction of the wave amplitude decrease.

Using standard boundary conditions at the vacuum-plasma interface and the condition (\ref{eq7}) at the gratings, let us exclude all internal waves and leave only equations which allow connecting the amplitudes of the reflected ($H_{2,0}$ and $H_{2,1}$) and transmitted ($H_{9,0}$ and $H_{9,1}$) wave harmonics with the incident wave amplitude $H_{1,0}$. These equations are the following: 
\begin{equation}\label{eq8}
\begin{array}{rrcrrl}
  H_{9,0} & +\alpha_{12}H_{2,0} &|&  & +\delta_1\alpha_{14} H_{2,1}&=\beta_1H_{1,0} \\
    & \alpha_{22}H_{2,0} &|& +\delta_1H_{9,1} & +\delta_1\alpha_{24}H_{2,1} &=\beta_2H_{1,0} \\
    -----&-----&|&-----&-----&\\
    & \delta_2\alpha_{32}H_{2,0} &|& +H_{9,1} & +\alpha_{34}H_{2,1} & =\delta_2\beta_3H_{1,0}\\
  \delta_2\alpha_{41}H_{9,0} & +\delta_2\alpha_{42}H_{2,0} &|&   & +e^{2q_1d_g}D_0H_{2,1} & =\delta_2\beta_4H_{1,0}\\
\end{array}
\end{equation}
The following notations are introduced here:
\[\delta_1={i\mu_1\over 2q_0},\,\,\delta_2={\mu_1\over 2q_1},\,\, q_1=\sqrt{(q_\xi+q_g)^2-\Omega^2}.\]
The function $D_0$ is determined by Eq.~(\ref{eq4}). The coefficients $\alpha_{ik}$ and $\beta_i$ do not vanish at $\mu_1=0$.

The matrix of coefficients in Eq.~(\ref{eq8}) has a block structure: two $2\times2$ blocks which do not contain the parameter $\mu_1$ are placed diagonally, and two anti-diagonal blocks which are proportional to $\mu_1$. Therefore for $\mu_1=0$ the system (\ref{eq8}) is split into two independent systems for propagating (with index $0$) and evanescent (with index $1$) harmonics. For $\mu_1=0$ a non-trivial solution for the propagated harmonics is possible only in the presence of the external wave $H_{1,0}$ (coefficient $\alpha_{22}\neq0$ always). However a non-trivial solution for the evanescent waves exists even when the external wave is absent. This solution exists if the determinant of the lower $2\times2$ block is equal to zero, or:
\begin{equation}\label{eq9}
D_0=0.
\end{equation}
The condition (\ref{eq9}) of course coincides with the dispersion equation (\ref{eq4}).

For small but finite $\mu_1\ll1$  the system (\ref{eq8}) determinant does not vanish at the roots of the dispersion equation (\ref{eq8}) but becomes small, $\sim\mu_1^2$. It means that at a frequency which is close to one of the surface waves resonator eigenfrequencies, the transmission coefficient defined as $|H_{9,0}/H_{1,0}|$ can be anomalously large.

Considering $\delta_1$ and $\delta_2$ as small, one can easily derive from Eq.~(\ref{eq8}) the following expression for the amplitude $H_{9,0}$ of the wave transmitted through the GPG system:
\begin{equation}\label{eq10}
H_{9,0}=H_{1,0}\left(A-\delta_1\delta_2{R \over
e^{2q_1d_g}D_0+\delta_1\delta_2P}\right),
\end{equation}
where $A$, $P$ and $R$ are certain combinations of the system (\ref{eq8}) coefficients $\alpha_{ik}$, $\beta_i$.

\section{Optically thick plasma layer}

The expressions for the Eq.~(\ref{eq10}) coefficients become essentially simpler when the plasma layer is thick, which is the case of the most interest. The coefficient $A$ determines the amplitude of the wave transmitted through the plasma  when the gratings are absent. This coefficient is exponentially small, $A\sim
e^{-\kappa_0d_p}\ll1$. Here $\kappa_0=\sqrt{q_\xi^2+1-\Omega^2}$ is the wave decay decrement in the plasma. When $\exp(-\kappa_0d_p)\ll1$, the expressions for the coefficients $R$ and $P$ can be written as:
\[R\simeq -4{\kappa_1^2\over q_1^2\varepsilon_p^2}B_+^2B_-^2e^{2\kappa_1d_p}e^{2i\beta}\cos^2\beta,\]
\[P\simeq -4{\kappa_1\over q_1|\varepsilon_p|}B_+B_-e^{\kappa_1d_p}e^{i\beta}\cos\beta,\]
where
\[B_\pm={1\over2}\left(1\pm{q_1|\varepsilon_p|\over\kappa_1}\right),\,\,
\kappa_1=\sqrt{(q_\xi+q_g)^2+1-\Omega^2}\] and
\[\beta=q_0d_g+\arctan{\kappa_0\over q_0|\varepsilon_p|}.\]

Using the functions $B_\pm$, expression (\ref{eq4}) for $D_0$ can be rewritten as 
\begin{equation}\label{eq10a}
D_0=B_-^2e^{\kappa_1d_p}-B_+^2e^{-\kappa_1d_p}.
\end{equation}

For $\mu_1\ll1$ the amplitude $H_{9,0}$ is maximal when the wave frequency is close to one of the resonator eigenfrequencies $\Omega_{res}$ which are defined by the dispersion equation roots, $D_0(\Omega_{res})=0$. The condition $\exp(-\kappa_0d_p)\ll1$ means that also $\exp(-\kappa_1d_p)\ll1$ and that the dispersion equation roots are exponentially close to the roots of the equation $B_-(\Omega)=0$. Therefore, one may put $q_1|\varepsilon_p|/\kappa_1\simeq1$ everywhere, except for $B_-$, and write the expression for the amplitude $H_{9,0}$ in the form:
\begin{equation}\label{eq10b}
H_{9,0}\simeq {i\mu_1^2\over
q_0q_1}{B_-^2e^{\kappa_1d_p}e^{-2q_1d_g}e^{2i\beta}\cos^2\beta\over
\left[\left(B_-^2-e^{-2\kappa_1d_p}\right)-i{\mu_1^2\over
q_0q_1}B_-e^{-2q_1d_g}e^{i\beta}\cos\beta\right]}H_{1,0}.
\end{equation}
Near the resonator eigenfrequency, which  is now determined by the equation  
$B_-^2=\exp(-2\kappa_1d_p)$,
Eq.~(\ref{eq10b}) can be presented as follows:
\begin{eqnarray}\label{eq11}
H_{9,0}\simeq {i\mu_1^2\over q_0q_1}{B_-(\Omega_{res})e^{\kappa_1d_p}e^{-2q_1d_g}e^{2i\beta}\cos^2\beta\over \left[2{dB_-\over d\Omega_{res}^2}\left(\Omega^2-\Omega_{res}^2\right)-i{\mu_1^2\over q_0q_1}e^{-2q_1d_g}e^{i\beta}\cos\beta\right]}H_{1,0}=\nonumber\\
\pm i\varepsilon_{eff}^2{e^{-2q_1d_g}e^{2i\beta}\cos^2\beta\over
\left[\left(\Omega^2-\Omega_{res}^2\right)-i\varepsilon_{eff}^2e^{-2q_1d_g}e^{i\beta}\cos\beta\right]}H_{1,0},
\end{eqnarray}
where
\[\varepsilon_{eff}^2={\mu_1^2\over 2q_0q_1\left(dB_-/d\Omega_{res}^2\right)}\]
and the signs $\pm$ correspond to the two roots of the dispersion equation $B_-(\Omega_{res})=\pm \exp(-\kappa_1d_p)$.

It is easy to see that the module of expression (\ref{eq11}) is maximal at the frequency which is determined by the condition $\mRe
\left[\left(\Omega^2-\Omega_{res}^2\right)-i\varepsilon_{eff}^2
e^{i\beta}\cos\beta\right]=0$ and reaches unity. In other words the transmission coefficient of an electromagnetic wave through the GPG system reaches 100\% at frequencies which are close to the surface waves resonator eigenfrequencies, and {it does not depend} on the value of the coefficient $\mu_1$. This coefficient determines only the bandwidth of the resonant transparency. 

The independence of the upper limit of the transmission coefficient from the value of the coefficient $\mu_1$, if it is small enough, confirms the statement that a concrete shape of a grating does not play an essential part. Various gratings, one- or two-dimensional, differ one from another by the value of the coefficient $\mu_1$, which describes a transformation of the incident propagating wave into the evanescent mode that excites the resonator.

\begin{figure}[tbh]
\centering \scalebox{0.6}{\includegraphics{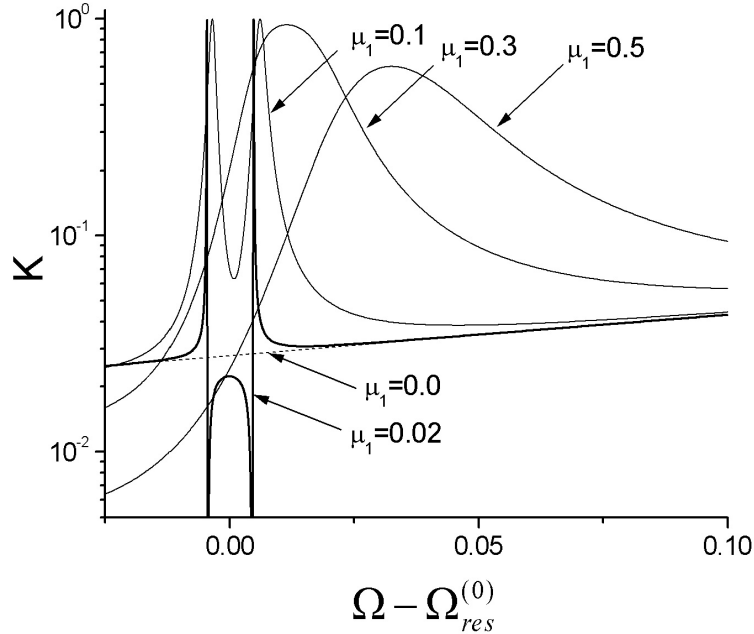}}
\caption{\label{Fig3}Transmission coefficient $K$ as a function of the normalized frequency $\Omega$. Layer thickness $d_p=3.0$.}
\end{figure}

These results of the approximate expression (\ref{eq11}) analysis are confirmed by a numerical solution of the system (\ref{eq8}). The dependencies of the transmission coefficient $K=|H_{9,0}/H_{1,0}|$ on the dimensionless frequency $\Omega$ for various values of the parameter $\mu_1$ are presented in Fig.~\ref{Fig3}. For convenience, the dependencies $K(\Omega-\Omega_{res}^{(0)})$ are shown. Here $\Omega_{res}^{(0)}$ are the roots of the dispersion equation (\ref{eq1}) for the isolated resonator. 
As it follows from the dependencies presented in Fig.~\ref{Fig3}, the plasmons excitation can lead not only to the transmission coefficient enhancement but to its suppression too. For small $\mu_1$ (thick curve $\mu_1=0.02$ in Fig.~\ref{Fig3}) the transmission coefficient in the close vicinity of the resonance frequencies can be much smaller than the transmission coefficient of the isolated plasma layer. In the model being used the plasmons excitation is the only mechanism which is responsible for the layer transparency enhancement. Therefore the film transparency suppression in some frequency regions cannot be used as an argument against the plasmon mechanism of the transparency enhancement (see \cite{Qing,Lezec}).

For small $\mu_1$ the dependence $K(\Omega)$ has the form of two reflection-symmetric Fano resonance profiles \cite{Fano}, that is caused by interference between the two terms in the right-hand side of Eq.~\ref{eq10}. The first term describes a non-resonant transmission of the wave through the plasma layer, the second one describes resonance transmission caused by the plasmons excitation \cite{Sarrazin2,Genet}. When $\mu_1$ increases, the Q-factor of the surface waves resonator decreases (see next section), which in turn leads to broadening of the resonance frequency bands. The overlapping of bandwidths of two resonances makes the role of non-resonance transmission negligibly small and the dependence $K(\Omega)$ acquires the typical form of two coupled resonators.

\section{Electromagnetic wave transmission through the GPG system as a resonator excitation}

The fact that the peak value of the transmission coefficient for small $\mu_1\ll1$ does not depend on the value of this parameter, has a simple explanation. The Q-factor of the resonator of the surface waves is determined by the energy flux out of the resonator. This flux is completely concentrated in the propagating waves. These waves arise at the diffraction gratings due to transformation of the evanescent waves, which are the eigenmodes of the resonator, into the propagating waves. The transformation coefficient is proportional to $\mu_1$. The resonator field amplitude at the grating is proportional to $\sim e^{-q_1d_g}$. Thus, the energy flux out of the resonator is proportional to $\sim\mu_1^2e^{-2q_1d_g}$ and the resonator Q-factor is inversely proportional to this value, $Q_{res}^{-1}\sim\mu_1^2e^{-2q_1d_g}$. Further, the resonator pumping is carried out by the non-propagating harmonic, which appears due to the external wave incidence at the first grating. This harmonic amplitude is proportional to  $\sim \mu_1 H_{1,0}$. The efficiency of an arbitrary resonator excitation by an external field $\Psi_{ext}(\zeta)$ is defined by the projection $\langle\Psi_{ext}\Psi_{res}^\ast\rangle$ of this field onto the resonator eigenmode field $\Psi_{res}(\zeta)$. In the case under consideration $\Psi_{ext}\sim \mu_1 H_{1,0}e^{-q_1\zeta}$,
$\Psi_{res}(\zeta)\sim e^{q_1(\zeta-d_g})$, i.~e. 
\[\langle\Psi_{ext}\Psi_{res}^\ast\rangle\sim\intop_0^{d_g}
\mu_1H_{1,0}e^{-q_1\zeta}e^{q_1(\zeta-d_g)}d\zeta\sim\mu_1
H_{1,0}e^{-q_1d_g}.\]

The steady-state amplitude $A_0$ of the oscillations in a resonator with a finite Q-factor is determined by the expression:
\[A_0\sim{\langle\Psi_{ext}\Psi_{res}^\ast\rangle\over \Omega^2-\Omega_{res}^2-i\Omega_{res}^2/Q_{res}},\]
where $\Omega$ and $\Omega_{res}$ are the external field frequency and the resonator eigenfrequency, respectively. In the case under consideration we obtain:
\[A_0=H_{7,1}\sim H_{1,0}{\mu_1 e^{-q_1d_g}\over \Omega^2-\Omega_{res}^2-ia\Omega_{res}^2\mu_1^2e^{-2q_1d_g}},\]
where the coefficient $a\sim 1$.

The field amplitude $H_{9,0}$ at the system output is determined by the product of the eigenmode amplitude at the grating surface,  $A_0e^{-q_1d_g}$, and the coefficient $\mu_1$ of the transformation of the evanescent wave  into the propagating wave:
\begin{equation}\label{eq12}
H_{9,0}\sim H_{1,0}{\mu_1^2 e^{-2q_1d_g}\over
\Omega^2-\Omega_{res}^2-ia\Omega_{res}^2\mu_1^2e^{-2q_1d_g}}.
\end{equation}

Eq.~(\ref{eq12}) has just the same structure as Eq.~(\ref{eq11}). The peak value of the transmission coefficient, as it follows from Eq.~(\ref{eq12}), is about unity and does not depend on the parameter $\mu_1$. Note, that in the estimations presented above of the resonator Q-factor and the resonator pumping efficiency the plasma thickness does not appear. It means that nominally an arbitrary thick metal film (plasma layer) can be made transparent using the construction under consideration.

The surface wave resonator properties are changed when diffraction gratings are placed near the plasma layer. Not only a finite Q-factor appears, as it was shown above, but the resonator eigenfrequencies are shifted as well. The frequency shift can be determined in the following way.

The dispersion equation Eq.~(\ref{eq4}) follows from standard boundary conditions at a vacuum-plasma interface. At the same time, from two possible solutions of the Maxwell equations in vacuum, only such a solution is chosen that decreases exponentially upon leaving the plasma boundary. This condition can be written as a Sommerfeld's condition analogue for evanescent waves:
\begin{eqnarray}\label{eq13}
\left.dH/d\zeta\right|_{d_g-0}-q_1\left.H\right|_{d_g-0}=0,\nonumber\\
\left.dH/
d\zeta\right|_{d_g+d_p+0}+q_1\left.H\right|_{d_g+d_p+0}=0.
\end{eqnarray}

In the presence of the gratings the conditions (\ref{eq13}) are no longer satisfied.
The ``medium'' at the left and at the right from the plasma boundaries  can now be characterized by the surface impedance $Z$ and the boundary conditions analogous to Eq.~(\ref{eq13}) take the form:
\begin{eqnarray}\label{eq14}
\left.dH/ d\zeta\right|_{d_g-0}+i\Omega Z\left.H\right|_{d_g-0}=0,\nonumber\\
\left.dH/ d\zeta\right|_{d_g+d_p+0}-i\Omega
Z\left.H\right|_{d_g+d_p+0}=0.
\end{eqnarray}

\begin{figure}[htb]
\centering \scalebox{0.6}{\includegraphics{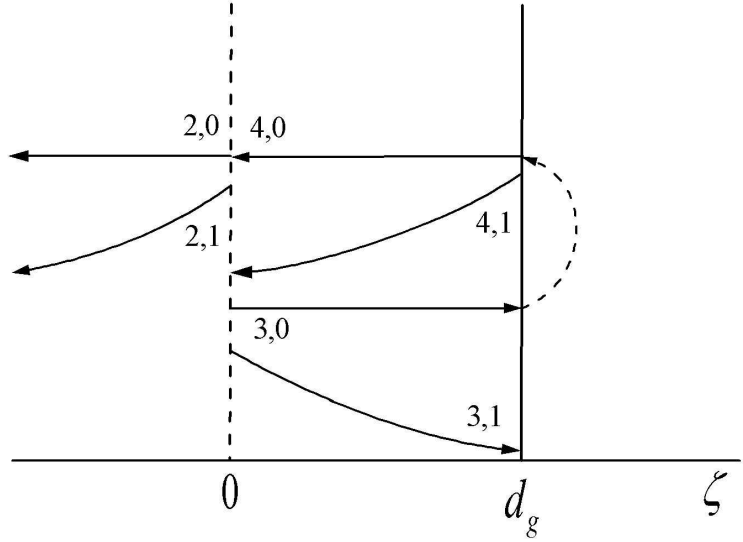}}
\caption{\label{Fig4}Resonator boundary impedance calculation.}
\end{figure}

Let us calculate the impedance considering, for example, the left plasma boundary. As it is represented in Fig.~(\ref{Fig4}), the wave $(4,1)$ which ``leaks'' from the plasma is scattered at the grating and additional waves appear. Neglecting the wave $(3,0)$ transmission through the plasma layer, it is possible to consider that this wave is almost reflected from the boundary and is transformed into the wave $(4,0)$. Using Eq.~(\ref{eq7}), one can connect the amplitudes of the waves $(4,1)$ and $(3,1)$ at the plasma boundary:
\begin{equation}\label{eq15}
H_{3,1}={i\over 4}{\mu_1^2\over
q_0q_1}\left(1+e^{i\alpha+2iq_0d_g}\right)e^{-2q_1d_g}H_{4,1},
\end{equation}
where $\alpha$ is the phase shift under the wave $(3,0)$ reflection.

At the plasma boundary the total magnetic field $H_1$ of a spatial harmonic, which is proportional to $\sim\exp[i(q_\xi+q_g)\xi]$, is equal to $H_{3,1}+H_{4,1}$. The tangential component of the electric field of this harmonic, $E_{\xi1}$, is equal to $-(i/\Omega)q_1(H_{4,1}-H_{3,1})$. Thus, as it follows from the impedance definition $Z=-E_{\xi1}/H_1$ (the surface normal is directed oppositely to the $\xi$-axis) and from Eq.~(\ref{eq15}):
\begin{equation}\label{eq16}
Z={i\over\Omega}q_1\left[1-i{\mu_1^2\over
q_0q_1}e^{iq_0d_g+i\alpha/2}e^{-2q_1d_g}\cos(q_0d_g+\alpha/2)\right].
\end{equation}
We put here $\mu_1^2\ll1$.

The impedance (\ref{eq16}) is a complex quantity. The imaginary part of the impedance is responsible for the eigenfrequency shift and the real part describes the resonator loss. Since the impedance imaginary parts of the ``medium'' with and without grating differ  from one another by the value 
\[\mIm(\delta Z)\sim {\mu_1^2\over q_0q_1}e^{-2q_1d_g}\sin(2q_0d_g+\alpha),\]
the resonator eigenfrequency shift is proportional to $\sim\mu_1^2$ for small $\mu_1$.

If the impedance of the ``medium'' surrounding the plasma layer is known, the resonator properties can be defined more precisely than it was done before. 

Using the boundary conditions at the vacuum-plasma interface and the conditions (\ref{eq14}), which can be rewritten in the form:
\begin{eqnarray}\label{eq17}
\left.dH/d\zeta\right|_{d_g-0}+q_\ast\left.H\right|_{d_g-0}=0,\nonumber\\
\left.dH/
d\zeta\right|_{d_g+d_p+0}-q_\ast\left.H\right|_{d_g+d_p+0}=0,
\end{eqnarray}
one can derive the following dispersion equation: 
\begin{eqnarray}\label{eq18}
D=\left[B_+(q_1+q_\ast)-B_-(q_1-q_\ast)\right]\left[B_-(q_1-q_\ast)-B_+(q_1+q_\ast)\right]e^{-\kappa_1d_p}-\nonumber\\
\left[B_-(q_1+q_\ast)-B_+(q_1-q_\ast)\right]\left[B_+(q_1-q_\ast)-B_-(q_1+q_\ast)\right]e^{\kappa_1d_p}=0.
\end{eqnarray}
Here
\[q_\ast=-iZ\Omega=q_1-i{\mu_1^2\over q_0}e^{-2q_1d_g}e^{i\beta_\ast}\cos\beta_\ast,\hspace{1mm}\beta_\ast=q_0d_g+\alpha/2.\]
Assuming that the plasma layer is optically thick, $\kappa_1d_p\gg1$, and $\mu_1^2\ll1$, it is possible to put $B_+\simeq 1$ in Eq.~(\ref{eq18}) and, keeping terms of order of at most $\mu_1^2$, Eq.~(\ref{eq18}) can be presented as follows:
\begin{equation}\label{eq19}
D\simeq
4q_1^2e^{\kappa_1d_p}\left[\left(B_-^2-e^{-2\kappa_1d_p}\right)-i{\mu_1^2\over
q_0q_1}B_-e^{-2q_1d_g}e^{i\beta_\ast}\cos\beta_\ast\right].
\end{equation}
The expression in the square brackets coincides exactly with the denominator in Eq.~(\ref{eq10b}). 

The dependence of the resonator eigenfrequencies on the grating transparency, which is characterized by the parameter $\mu_0$ in Eq.~(\ref{eq5}), can be obtained in an easier way putting $\mu_1=0$. Replacing the second equation in (\ref{eq7}) by the boundary condition
\begin{equation}\label{eq20}
\left.{dH_n\over d\zeta}\right|_{+0}-\left.{dH_n\over
d\zeta}\right|_{-0}+\mu_0H_n(0)=0,
\end{equation}
it is easy to obtain the following dispersion equation:
\begin{eqnarray}\label{eq21}
\left[{\mu_0\over 2q_1}e^{-q_1d_g}B_++\left(1-{\mu_0\over 2q_1}\right)e^{q_1d_g}B_-\right]^2e^{\kappa_1d_p}-\nonumber\\
\left[{\mu_0\over2
q_1}e^{-q_1d_g}B_-+\left(1-{\mu_0\over2
q_1}\right)e^{q_1d_g}B_+\right]^2e^{-\kappa_1d_p}=0.
\end{eqnarray}
When the plasma layer is rather thick, $\exp(\kappa_1 d_p)\gg1$, then the roots of the dispersion equation (\ref{eq21}) are close to the roots of the first preexponential factor and Eq.~(\ref{eq21}) can be simplified:
\begin{equation}\label{eq22}
{\mu_0\over2
q_1}e^{-q_1d_g}B_++\left(1-{\mu_0\over2
q_1}\right)e^{q_1d_g}B_-=0.
\end{equation}

Let us restrict ourselves to the case when the gratings are situated close to the plasma surfaces \footnote{When the gratings are situated rather far from the plasma surfaces so that $q_1d_g\geq1$, then an extra pair of roots appears. These additional roots do not disappear for $\mu_0\rightarrow\infty$ and in some range of $\mu_0$ variation, four roots can exist simultaneously. The transmission coefficient $K$ is extremal in the roots locations. In the considered frequency range and for $q_1d_g\geq1$ one more mechanism exists, which is not connected with the surface waves resonator and also leads to the plasma layer transparency. The overall picture of the wave transmission is more complex in this case and will not be considered here.}, when
$\exp(q_1d_g)\simeq1$. Using the definitions of the functions $B_\pm$, let us write Eq.~(\ref{eq22}) in the form:
\begin{equation}\label{eq23}
{q_1|\varepsilon_p|\over\kappa_1}\left(1-{\mu_0\over
q_1}\right)=1.
\end{equation}

\begin{figure}[tbh]
\centering \scalebox{0.6}{\includegraphics{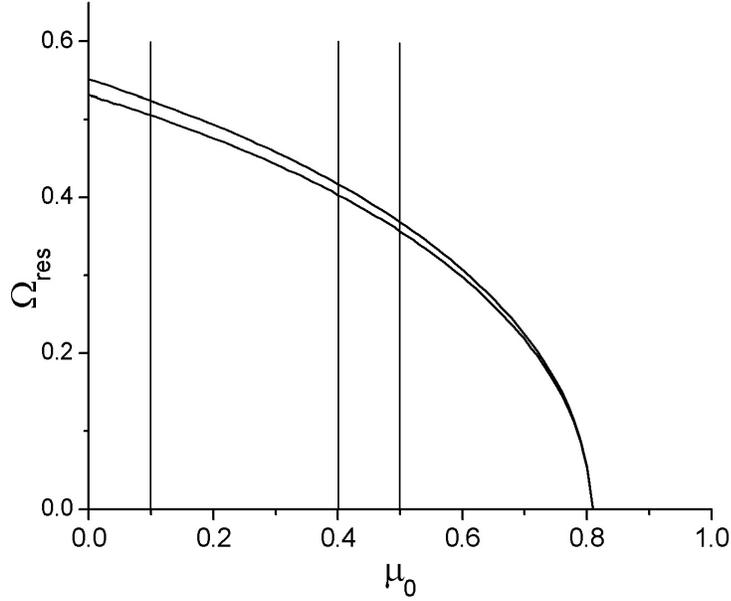}}
\caption{\label{Fig5}Dependence of the resonator eigenfrequencies on the gratings transparency (parameter $\mu_0$).}
\end{figure}

It is easy to see that the roots $\Omega_{res}$ of the dispersion equation (\ref{eq23}) decrease with the growth of $\mu_0$ and reach zero for $\mu_0=q_\xi+q_g$. For $\mu_0>q_\xi+q_g$ Eq.~(\ref{eq23}) has no real roots. This dependence $\Omega_{res}(\mu_0)$, which has been obtained numerically from Eq.~(\ref{eq21}), is presented in Fig.~\ref{Fig5}.

\begin{figure}[tbh]
\centering \scalebox{0.6}{\includegraphics{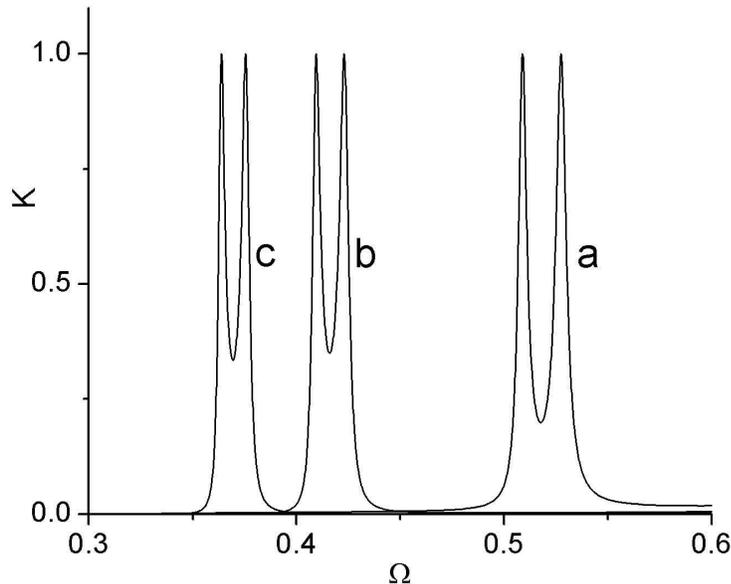}}
\caption{\label{Fig6}Dependence of the transmission coefficient $K$ on the wave frequency for several values of $\mu_0$: (a) $\mu_0=0.1$, (b) $\mu_0=0.4$, (c) $\mu_0=0.5$.}
\end{figure}

The resonator considered above  is a resonator of evanescent waves, therefore for $\mu_1=0$ its Q-factor is infinitely large. For $\mu_1\neq0$ the transformation of the evanescent waves into the propagating ones makes the resonator Q-factor finite and allows the excitation of the resonator by an external propagating wave.
As in the case for $\mu_0=0$, the GPG system transparency can reach 100\% when the incident wave frequency is close to one of the resonator eigenfrequencies $\Omega_{res}$. The dependence $\Omega_{res}(\mu_0)$ described above is fully confirmed by the numerical solution of the system (\ref{eq8}). For example the dependencies $K(\Omega)$ for several values of $\mu_0$ marked by the vertical lines in Fig.~\ref{Fig5}, are presented in Fig.~\ref{Fig6}.

Fig.~\ref{Fig6} also demonstrates  that the eigenfrequencies of the plasma layers which are either isolated ($\mu_0=0$) or surrounded by gratings  can differ greatly and this difference cannot be treated as an argument against the plasmon mechanism of the transparency enhancement, as it was done, for example, in Ref.~\cite{Krishnan}.

In closing this section let us note that the above-mentioned equivalence between electromagnetic wave transmission through a plasma layer and excitation of two coupled identical resonators allows one to determine how one or another factors unaccounted for in this paper affect the result. Thus, for example, dissipative losses into the plasma layer can be taken into account in the following simple way. Making use of the analogy between a resonator and an oscillator, it is easy to see that the coupling between two oscillators (between two resonators, or, in other words, between the electromagnetic fields on either side of the layer) is destroyed when the generalized coupling coefficient $c_{coupl}$ is small as compared with the generalized friction coefficient $c_{fric}$, 
\begin{equation}\label{eq24}
c_{coupl}\ll c_{fric}.
\end{equation}

In the case considered here, the coupling coefficient is determined by overlapping of the evanescent fields of eigenmodes of two resonators , $c_{coupl}\propto d_pe^{-\kappa_1d_p}$. The friction coefficient is determined by dissipation in the layer, $c_{fric}\propto d_p {\rm Im}\,\varepsilon_p$. Thus, the resonant transmission through the layer can be observed when the following condition is satisfied:
\begin{equation}\label{eq25}
{\rm Im}\,\varepsilon_p e^{\kappa_1d_p}\ll1.
\end{equation}
Note that just this condition restricts the superresolution of the "perfect lens" \cite{Neto-Vesperinas}.

\section{Conclusion}

The merit of the suggested model  is that all other possible mechanisms (waveguide connection between the film surfaces through small holes, action of holes as subwavelength cavities for the evanescent waves, interference of diffracted evanescent waves, {\it etc.}) of the electromagnetic wave transmission through a perforated or corrugated metal film are absent {\it a priori}. Excitation of the surface waves resonator is the only mechanism in this model which is responsible for the electromagnetic wave transmission through a metal film. The knowledge of the resonator properties is sufficient for determining how the GPG system parameters affect the frequency and incidence angle for which the film transparency is maximal.

The resonator mechanism can be considered as one of possible competitive mechanisms of light transmission through a perforated metal film. A periodically inhomogeneous film surface can be considered as a diffraction grating situated directly at the film surface, and holes can be considered as an additional channel of connection between the resonators at the two opposite film surfaces. The increase of the coupling between the resonators decreases the destuctive influence of dissipation on the wave transmission through the film. The analysis of the dependence of the resonator eigenfrequencies on the system parameters shows that the eigenfrequencies of a smooth film and the grating-plasma-grating system can be significantly different. Therefore even a strong deviation of the frequency, which corresponds to the transparency peak, from the plasma layer (metal film) eigenfrequency, cannot be unambiguously treated as a sign of another, different from the resonator, mechanism of electromagnetic wave transmission through the metal film.

\section{Acknowledgments}
I am grateful to Professors Y. Ben-Aryeh and J. Felsteiner for discussions that stimulated this work. The research has been supported by the Center for Absorption in Science of the Ministry of Immigrant Absorption of Israel.

\end{document}